\documentclass{bcp80}
\usepackage{hyperref}
\usepackage{colordvi}
\usepackage{graphicx}
\usepackage{here}
\renewcommand{\Red}[1]{{#1}}

\def\LaTeX{L\kern -.36em\raise .3ex\hbox{\sc a}\kern -.15em T\kern -.1667em%
\lower .7ex\hbox{E}\kern -.125em X}

\begin{document}

\keywords{evolutionary game theory, finite populations, asymmetric conflicts}
\mathclass{Primary 91A22; 
Secondary 
 60J20
, 92D50
, 92D15
.}
%
\abbrevauthors{Jens Christian Claussen}


\abbrevtitle{Coevolutionary dynamics in finite and infinite populations}

\title{%
Discrete stochastic processes, replicator 
and
\\ Fokker-Planck equations of coevolutionary 
\\
dynamics in finite and infinite populations
}

\author{Jens Christian Claussen}
\address{Institute of Theoretical Physics and Astrophysics, 
Christian-Albrecht University Kiel\\
Leibnizstr.\ 15, 24098 Kiel, Germany\\
E-mail: claussen@theo-physik.uni-kiel.de}

\maketitlebcp

\abstract{Finite-size fluctuations in coevolutionary dynamics arise 
in models of biological as well as of social and economic systems. 
This brief tutorial review 
surveys a systematic approach starting from a
stochastic process discrete both in time and state.
The limit $N\to \infty$ of an infinite population can be considered
explicitly, generally leading to a replicator-type 
equation in zero order, and to a 
Fokker-Planck-type equation in first order in $1/\sqrt[]{N}$.
Consequences and relations to some previous approaches are outlined.
}


\section{Introduction.}
Evolution is a biological process ubiquituosly taking place,
acting on on several temporal, spatial, and taxonomic scales.
Biological organisms are coded by their genetic sequence,
and their ability to survive, following Darwin, is largely
determined by their genes. 
However, even the relationship between genotype, phenotype
and a fitness (e.g., defined by the reproduction rate)
by no means is simple, 
as it maps an 
enormously high-dimensional space 
(the state space of all possible genomes)
to a much lower-dimensional phenotype space,
and finally to a one-dimensional space of
fitness values.

The extinction of species, as observed in the fossile
record, follows a complicated behaviour 
 \cite{Eld72}
including intermittency-like 
long periods of stasis,
the so-called ``punctuated equilibria''.
It is still at debate whether 
external or cosmic influences 
account for these, or 
whether a purely dynamic mechanism
of the evolutionary and ecological
dynamics can produce this stylized fact.
In this direction, the Bak-Sneppen model
\cite{Bak93} is a pioneering 
minimal model
for the extinction of species, 
which however is difficult to relate
to the biological scenario.
On the one hand, it provides an intermediate modeling level,
where biological observations can be explained to some extent,
but
on the other hand,
a mathematical treatment is still possible
\cite{Ban05,Pis97}.
On an individual-based level,
the Tangled Nature model
\cite{Jen04}
approaches closer to biology.
While the dynamics at large 
will deserve some further decades of 
research, 
simplified situations of coevolutionary
dynamics,
restricting to a finite number of species
can be studied by formal models 
of evolutionary processes, 
and can be solved analytically in many cases
to allow for an exemplaric insight.
In this brief tutorial review, coevolutionary
dynamics in finite populations
is formulated within the framework
of evolutionary game theory, 
providing a convenient common mathematical
framework for biological, social and economical
evolutionary processes of strategies, or
 genetic types.

\subsection{Game theory: The strategy of conflict.}
Game theory was brought into play 
by von Neumann and Morgenstern
\cite{Neu53}
as a minimal model to explain and predict the
behaviour of humans in strategic
situations,
be them military or economic,
assuming fully rational behaviour.
In game theory, 
more specific in two-player normal form games,
agents 
$1 \ldots N$
``play'' a strategy,
out of a
finite set of possible pure strategies.
They interact with a partner
also playing one of those 
stragegies, and both receive a (real valued) payoff according to
a so-called payoff matrix. 
This is best illustrated with an
example.
The most paradigmatic conflict situation described in game theory
is the Prisoner's Dilemma.
It is
defined by the payoff matrix
\begin{eqnarray}
\pi_{ij}=
\left(
\begin{array}{cc}
3 & 0 \\
5 & 1
\end{array}
\right).
\end{eqnarray}
This is to be read as follows. 
The game is defined between two players,
who can adopt two possible strategies:
to cooperate (C)
or to defect (D).
The players act 
in
parallel and are not informed about the
opponent's move.
Each player P (row player) playing strategy $i$ 
then receives the payoff $\pi_{ij}$ 
when meeting a player using strategy $j$ 
(column player, O) as opponent.
Obviously, here the conflict situation is symmetric,
therefore
the payoff matrix of the opponent's payoffs is
given by the transpose $\pi_{ij}^{\rm T}$.
Often both payoff matrices are combined to
\begin{eqnarray}
(\pi_{ij}^{\rm P},\pi_{ij}^{\rm O})=
\left(
\begin{array}{cc}
3,3 & 0,5 \\
5,0 & 1,1
\end{array}
\right).
\end{eqnarray}
In symmetric conflicts, thereby only redundant information is
added. 
In asymmetric conflicts, 
for obvious reasons often called {\sl bimatrix games},
the opponent's payoffs, in general, can be different.
We will analyze such cases in Section \ref{Cla:Sec:Asym}.

In all cases above, the players play their respective strategy with
probability one, i.e., they play pure strategies.
A straightforward generalization is that players
--- in a random and uncorrelated manner ---
can use {\sl mixed strategies}, 
or {\sl strategy profiles} $s_n$,
i.e., possess a 
strategy vector 
(with components $(s_n)_i$
summing up to one, thus are elements of a simplex $S_n$),
and gain payoffs which
are the corresponding linear combination 
\begin{eqnarray}
u_n(s_n,s_m) = \sum_{ij} (s_n)_i  \pi_{ij} (s_m)_j
\end{eqnarray}
for playing against one opponent (profile) $s_n$, and
\begin{eqnarray}
u_n(s_n,\{s_m\}) = \sum_{ijm} (s_n)_i  \pi_{ij} (s_m)_j
\end{eqnarray}
against a set of opponents indexed by $m$.
Now, can mixed strategies exist that are
successfull against all hypothetical sets 
of opponents?
This brings us to the central concept of a Nash equilibrium \cite{Nas51}:
A Nash equililbrium is a
mixed strategy,
in which
no single agent can improve its payoff by
solely changing its own strategy.

\subsection{Nash equilibria.}
A {\sl strict Nash equililbrium (resp., Nash equililbrium)} is defined as a 
strategy profile $s^*=(s^*_1,\ldots s^*_N)$ for which
each agent's strategy $s_n$ is a best
(resp., best or equal) response to the strategies of the other
players $s^*_{-n}$,
i.e.\
\begin{eqnarray}
\forall_{n} \forall_{s \in S_n, s \neq s^*_n} 
\;\;\;\;
u_n(s^*_n,s^*_{-n}) \geq u_n(s,s^*_{-n})
\end{eqnarray}
where $>$ instead of $\geq$ must hold for a strict Nash equilibrium.
\\ 
Hereby $s^*_{-n}:=(s_1, \ldots s_{n-1},s_{n+1}, \ldots s_{N})$
is called a strategy profile of the 
co-players
(formally a set of $N-1$ 
 profiles, usually it is
 understood that playing against all co-players 
 linearly sums the payoffs received playing 
 with each of them; 
 so one can define the co-profile 
 $(s^*_{-n})_k:= (N-1)^{-1} \sum_{i \neq n} (s_i)_k$
 as a linearly averaged profile of the co-players;
or 
define 
$u_n( . ,s^*_{-n})$ as the sum of payoffs against 
the profile of each co-player).
\\ \indent
Nash's theorem \cite{Nas51} ensures that 
a normal-form game (as defined above) for a finite number of
strategies and 
a finite number of players 
always posesses a Nash equilibrium.
However, it can be degenerate or 
a mixed strategy {\sl (mixed Nash equlibrium)}.
%
In the above
 Prisoner's Dilemma,
$((1,0),(1,0))$, 
i.e.\ both players playing always 
``defect'', is a Nash equlibrium;
hence two memory-less agents playing it for one round
have no incentive to cooperate.
Conversely,
the Stag-Hunt game
$
\pi_{ij}=\left(\begin{array}{cc} 5 & 0 \\3 & 3\end{array}\right)
$
has
 two symmetric;
$((1,0),(1,0))$,
and $((0,1),(0,1))$
 the Hawk-Dove game
$
\pi_{ij}= \left(\begin{array}{cc} (V-C)/2 & V \\ 0 & V/2 \end{array} \right)
$
with $C>V$ has two nonsymmetric 
Nash equilibria,
$((1,0),(0,1))$ and $((0,1),(1,0))$,
the latter is an example of a population where 
both strategies are present.

\subsection{Evolutionary Game Theory
and evolutionarily stable strategies.
}
Decades after game theory was invented, 
Maynard Smith and Price 
\cite{May73}
were the first to utilize its approaches to the
understanding of biological conflict situations
among whose the emergence of cooperation
\cite{Axe84} 
among animals and humans still is a continuously 
active field.
In the dynamical picture of evolutionary game
theory,
the concept of the Nash equililbrium 
has its counterpart in the 
{\sl evolutionarily stable strategies} (ESS),
which are defined as a population 
in which a single mutant (changing to any of the
possible strategies or genotypes)
cannot invade the population.
In infinite populations, traditionally described
within the framework of replicator equations
\cite{Tay78,Hof79,Zee80},
ESS appear as stable fixed points
(see \cite{Hof98} for a systematic treatment).
In finite populations, however,  
this concept has to be refined
\cite{Now04,Tay04,Nei04,Wil04,Tra06b}.
%
For a more detailed introduction into the field
of evolutionary game theory,
and its recent development, see
the classical textbooks 
\cite{Hof84,Hof98}
and recent reviews \cite{Sza07,Mie07}.
\\[0mm]

\clearpage

\section{Microscopic interactions:
 Game theory based on particle collision models.}
The description of social agent behaviour 
as interaction, or collision, of particles 
has been studied by Helbing establishing a quite
general framework
\cite{Hel92a,Hel92b,Hel96},
which at that time had not further been taken up within
evolutionary game theory or evolutionary dynamics.

One main limiting assumption of the
mean-field type description discussed in the next sections
is that any spatial organization can be neglected
in a first order approximation.
This approximation, as well as neglecting age structure and 
time delay \cite{Alb04}, however is not warranted in general.

\subsection{Patchy ecosystems.}
To describe the full dynamics of ecological and evolutionary processes,
it can be necessary to split up the population into 
parts, or patches, and to investigate an intermediate level of
subdivided populations \cite{Che03} or
metapopulation dynamics \cite{Har02}.
In general, a rich variety of dynamical scenarios can emerge; 
thus simplified models on lattices and graphs have been
investigated widely.


\subsection{Spatial models.}
Life typically is organized, to a very rough approximation,
as a covering of the surface of earth, in competition
for sunlight, solid ground, or hunting territory.
So it is natural to investigate evolutionary dynamics 
of individuals located in a two-dimensional space
\cite{Now92}, 
where collective phenomena can emerge \cite{Her94}.
Annother classical study by Lindgren and Nordahl
\cite{Lin94}
investigated the spatial Prisoner's 
Dilemma game with strategies of different memory lengths.
Spatial structure has been identified as one 
possible mechanism to promote cooperation \cite{Now92}.
In general, spatial game theory is capable of 
rich dynamical behaviour
\cite{Sza99,Sza04a,Sza05,Tra04},
as coarsening, segregation, and spiral waves.
Likewise, models for opinion dynamics have been studied,
as the Sznajd model
\cite{Szn00}.

\subsection{Evolutionary dynamics on graphs.}
The systematic understanding of evolutionary
dynamics on graphs is a still developing field.
An early investigation of coevolutionary dynamics 
(i.e., including frequency-dependence)
on graphs has been given by
Ebel and Bornholdt 
\cite{Ebe02}, 
investigating the iterated Prisoner's Dilemma on networks.
On the small-world architecture, Szab\'{o} et al.\
\cite{Sza04b} 
investigated the Rock-Paper-Scissors game dynamics.
In \cite{Lie05,Oht06},
fixation properties of evolutionary dynamics on graphs 
are studied, and special subgraphs have been
identified to enhance or suppress fixation.
For two strategies and non frequency-dependent fitness,
Antal, Redner and Sood \cite{Ant06}
have provided exact results for fixation 
for the case of degree-uncorrelated graphs.
Despite this significant progress \cite{Sza07}, 
a general theory of
coevolutionary dynamics on graphs
remains a formidable challenge.

\subsection{Unstructured population dynamics: 
Meanfield approach, or P\'{o}lya urn models.}
The approximation of an unstructured population 
implies that individuals are undistinguishable
(apart from their strategies or genotype), 
and individuals are chosen randomly from the population,
for death, reproduction, and competition.
Hence, the population can be viewed as a P\'{o}lya urn 
from which individuals are drawn for the stochastic process.
All evolutionary processes discussed
for finite populations 
in the remainder
belong to this class of processes.
\\[-10mm] 


%
\clearpage
\section{Microscopic evolutionary processes} 
\begin{figure}[htbp]
\begin{center}
\setlength{\unitlength}{1pt}
\begin{picture}(320,120)(-10,-10)
\put(0,20){\circle{10}}
\put(20,20){\circle{10}}
\put(40,20){\circle*{10}}
\put(60,20){\circle*{10}}
\put(80,20){\circle*{10}}
\put(100,20){\circle{10}}
\put(0,65){\vector(0,-1){30}}
\put(2.5,65){\vector(1,-2){15}}
\put(22.5,65){\vector(1,-2){15}}
\put(77.5,65){\vector(-1,-2){15}}
\put(80,65){\vector(0,-1){30}}
\put(100,65){\vector(0,-1){30}}
\put(0,80){\circle{10}}
\put(20,80){\circle*{10}}
\put(40,80){\circle*{10}}
\put(60,80){\circle*{10}}
\put(80,80){\circle*{10}}
\put(100,80){\circle{10}}
\put(200,-10){\circle{10}}
\put(200,35){\vector(0,-1){30}}
\put(200,50){\circle{10}}
\put(220,40){\circle{10}}
\put(220,85){\vector(0,-1){30}}
\put(220,100){\circle*{10}}
\put(240,10){\circle*{10}}
\put(240,55){\vector(0,-1){30}}
\put(240,70){\circle*{10}}
\put(260,20){\circle{10}}
\put(260,65){\vector(0,-1){30}}
\put(260,80){\circle*{10}}
\put(280,30){\circle*{10}}
\put(280,75){\vector(0,-1){30}}
\put(280,90){\circle*{10}}
\put(300,0){\circle*{10}}
\put(300,45){\vector(0,-1){30}}
\put(300,60){\circle{10}}
\end{picture}
\end{center}
\caption{Discrete stochastic evolution processes: 
Fisher-Wright process (left) 
as a typical synchroneous update process,
and an
overlapping generations process 
(right).
In the Moran process and Local update
the lifetime of the individuals 
however is nonidentical, due to the 
stochastic asynchroneous update.
For the transition probabilities 
see text.
The open and closed bullets can represent 
two different strategies (in a social system),
or two different alleles at a specific genetic
locus (in biology). The generalization to 
higher numbers of strategies or genetic types 
is straightforward.
\label{fig_fisherwrightmoran}}
\end{figure}
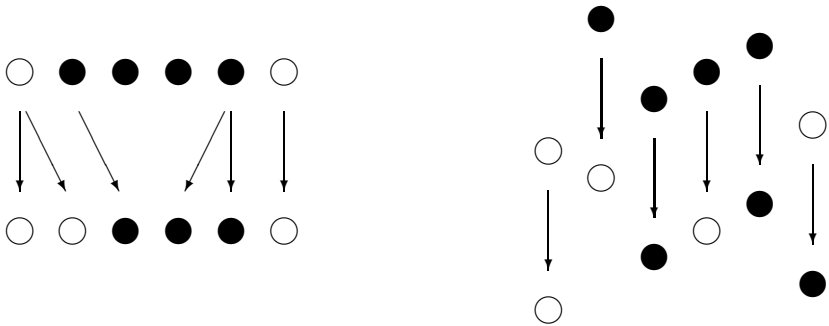

Evolutionary processes have been widely considered 
in population genetics.
For processes in discrete time, an important systematic distinction
has to be made between synchroneous 
or parallel \cite{Baa97} update processes,
as the Fisher-Wright process,
and processes with overlapping generations, where one
individual is replaced in each evolutionary step
(see Fig.\ \ref{fig_fisherwrightmoran}).


\subsection{Fisher-Wright process.} 
The process defined by Fisher
\cite{Fis30} and Wright \cite{Wri31}
in its original form is not frequency-dependent;
the fitness of the individuals does not depend on the state
of the population (given by the number of individuals in 
each of the strategies).
The Fisher-Wright process is defined as follows. In each time step, all
individuals reproduce
with probabilities proportional to their fitness,
until the same population size $N$ is reached.
%
An important case is given by {\sl neutral evolution},
where all individuals have identical fitness.
Hence, the discrete stochastic process 
that describes the time evolution 
resembles a random genetic drift, and
no Darwinian principle is incorporated.

Of course the Fisher-Wright process can be straightforwardly generalized
to the case where the fitness of strategies depends on
payoff gains obtained from game-theoretic interactions with
other individuals. 
The finite population case
has been considered recently by Taylor and Nowak
\cite{Tay06}.
Here the transition matrix of the Markov process
is by no means sparse; apart from the absorbing 
boundaries all matrix elements can be nonzero.


\subsection{Discrete stochastical processes for overlapping generations.}
The evolutionary processes 
of Moran or Local update type
provide us with a transition probability 
$T^{\pm}$ that the number of individuals $i$ (being in the first
strategy) increases or decreases by one, respectively.

For definiteness, we consider explicitly the
Moran evolution dynamics in arbitrary $2{\times}2$ games
\cite{Cla05}.
Given a finite population of N agents  (two types/strategies A and B) 
interacting via
 in a game with the payoff matrix
\begin{eqnarray}
P=\left( \begin{array}{cc} a & b \\ c & d \\ \end{array} \right).
\end{eqnarray}
In the frequency-dependent
Moran
 process,
Every agent interacts with a 
representative sample of the population,
yielding the payoffs
of $A$ and $B$ individuals 
as
\begin{eqnarray}
\pi^A(i) & = &
1 - w + w 
\frac{a(i-1)+b(N-i)}{N-1} \\
\pi^B(i) & = &
 1 - w + w 
\frac{c \, i+d(N-1-i)}{N-1},
\end{eqnarray}
$i$ is the number of $A$ individuals, 
and
 $1-w$ is a background fitness.

\subsection{Moran process.}
In its original form, also the Moran process
\cite{Mor62} 
is not frequency dependent. 
In this birth-death process, in each time step 
an individual is selected for reproduction, 
and subsequently a randomly selected individual dies.
Hereby 
it is ensured that
the total number of individuals $N$
remains
constant.

For two strategies, the Moran process therefore is
a one-dimensional Markov process with a finite 
number of states from $i=0,\ldots,N$.
Taking up the Moran approach, Nowak,
Taylor, Fudenberg and Sasaki \cite{Now04,Tay04}
investigated a frequency-dependent Moran process,
defined as follows:
In the frequency-dependent Moran process, 
selection for reproduction is 
proportional to the payoff compared to the average payoff,
$\pi^A(i)/\langle\pi\rangle$.
The probability per time step that a copy of an A agent 
is newborn then is
$p^{+}i/N$
(with $p^{+}$ as given in Table \ref{table1}).
It replaces a randomly chosen individual.
Hereby, the fitness is evaluated after each individual
took interaction according to a payoff matrix with 
all individuals in the population.

\subsection{Local update and imitation processes.}
An apparent limitation of the Moran process with
respect to biological situations is that
individuals in each update step have to compete 
with the whole population
(or, in real systems
with a representative sample) 
of the population.

A more realistic setting is given by local or
pairwise competetive interactions, which can be
called imitation dynamics \cite{Hof00}
(in the concept of social
strategies) 
or local update. 
These processes can be defined in a slightly different 
manner. 
In a non-symmetric definition of the (linear) local 
update process \cite{Tra05}, one individual is 
selected for reproduction, and the other for death.
Then the strategy of the reproducing individual
is changed (or kept) with probabilities 
that depend
linearly on the difference of the
payoffs of the two interacting individuals.
Hereby, as is usually assumed in evolutionary game theory,
the payoffs are evaluated by playing
(on a shorter time scale) with the 
whole population,
so these payoff differences still are evaluated 
in a global process. 

\begin{table}[bhpt]
\caption{Comparison of three microscopic update processes
and their respective differential equations 
for $x:=i/N$
obtained in the 
deterministic limit of $N\to\infty$
(see Sec.\ \ref{secLIMIT}).
\label{table1}}
\noindent
\begin{tabular}{|lll|}
\hline
Moran process&Local update&Fermi process\\
\hline
$\displaystyle
p^+=
\frac{1-w+w\pi_i^A}{1-w+w\langle\pi_i\rangle}
$&
$\displaystyle
p^+=
\frac{1}{2}+\frac{w}{2}
\frac{\pi_i^A-\pi_i^B}{\Delta \pi_{\rm max}}
$
&
$\displaystyle
p^+=
\frac{1}{1+{\rm e}^{-w (\pi_i^A-\pi_i^B)}}
$
\\
$\displaystyle
p^-=
\frac{1-w+w\pi_i^B}{1-w+w\langle\pi_i\rangle}
$&
$\displaystyle
p^-=
\frac{1}{2}+\frac{w}{2}
\frac{\pi_i^B-\pi_i^A}{\Delta \pi_{\rm max}}
$
&
$\displaystyle
p^-=
\frac{1}{1+{\rm e}^{-w (\pi_i^B-\pi_i^A)}}
$
\\
&&
\\
$0\leq w \leq 1$&
$0\leq w \leq 1$&
$0\leq w \leq \infty$
\\
\hline
 Transition probabilities
&&
\\for increase/decrease of $i$:
&&
\hspace*{-2em}
\\
$\displaystyle
T^{\pm}=p^{\pm}\frac{i}{N}\frac{N-i}{N}$
&
$\displaystyle
T^{\pm}=p^{\pm}\frac{i}{N}\frac{N-i}{N}$
&
$\displaystyle
T^{\pm}=p^{\pm}\frac{i}{N}\frac{N-i}{N}$
\\
\hline
In the $N\to\infty$ limit:
&&\\
\small
$ \displaystyle
\dot{x}=
\frac{w
x (1-x) (\pi^A_x
-\pi^B_x
)
}{1-w + w\langle \pi(x) \rangle}
$
&
\small
$
\displaystyle
 \dot{x}=
\frac{w
x (1-x) (\pi^A_x
-\pi^B_x
)
}{\Delta \pi_{\rm max}}
$
&
\small
$
\displaystyle
 \dot{x}=
x(1-x)
\tanh(\frac{w}{2}(\pi^A_x
-\pi^A_x
))
$
\\
{\sl adjusted replicator 
eq.\
}
\hspace*{-.7em}
&
{\sl 
ordinary
 replicator eq.\  
}
&
{\sl
replicator eq.\
after} \cite{Tra06c}
\\
\hline
\end{tabular}
\end{table}

\subsection{Nonlinear response: Local Fermi process.}
A variant 
\cite{Blu93,Sza98,Hau05}
of the Local update process arises naturally
when the game interaction payoff is one of many additive 
contributions to the reproductive fitness. 
In the case where the external contributions are
large, and consist of many degrees of freedom that
act like an external heatbath, 
the payoffs of the individuals appear as argument of Boltzmann
factors, where a parameter $w$ can be introduced
as for the other processes ($w\to 0$ corresponds to weak selection),
and here can be interpreted as a temperature.
This pairwise comparison process has been studied in detail in
\cite{Tra06c,Tra07}.
See Table \ref{table1} for the
 transition
probabilities 
$T^{\pm}$
of the Moran process, Local update, and
Fermi process.

\section{Broadening of distributions in finite populations.}
For the Moran process, the strategy distribution is generated only by the
 inherent stochasticity of the finite population.
At the borders, for $T_{0\to 1}$ and $T_{N\to N-1}$ 
we assume a small mutation rate $\mu$ \cite{Fud04}.
While in the 
infinite 
population case the population 
density is peaked at the deterministic trajectory in the
sense of a delta distribution, 
for a finite population it is broadened.
Fortunately, the Moran process allows for an analytical
treatment.
The stationary distribution for an arbitrary payoff matrix
can be expressed via Poch\-ham\-mer symbols, rising factorials
or gamma functions,
for the  general case of
$2{\times}2$ games
 including nonvanishing background fitness
\cite{Cla05}.
For illustration, let us consider 
the special cases 
of 
neutral evolution,
constant fitness,
an
`anticoordination game'
and the Prisoner's dilemma,
\begin{eqnarray}
\nonumber
\mbox{
~~~
$P_{\rm n} =\left( \begin{array}{cc} a & a \\ a & a \\ \end{array} \right)$,
~~~
$P_{\rm c} =\left( \begin{array}{cc} a & a \\ c & c \\ \end{array} \right)$,
~~~
$P_{\rm AC} =\left( \begin{array}{cc} 0 & 1 \\ 1 & 0 \\ \end{array} \right)$,
~~~
$P_{\rm PD} =\left( \begin{array}{cc} 3 & 0 \\ 5 & 1 \\ \end{array}
\right)$.
}
\end{eqnarray}

\subsubsection*{Internal 
(mixed)
 vs.\ external 
(pure) 
Nash equilibrium.}
For frequency dependent fitness and $w>0$, the game can 
have an internal Nash equilibrium or an
equilibrium in one of the absorbing states. 
A simple example with an internal Nash equilibrium 
is 
the  `anticoordination' game with $w=1$,
$
P=
\Big( \begin{array}{cc} 0 & 1 \\ 1 & 0 \\ \end{array} 
\Big).
$
%
The transition matrix here is
$
T_{i\rightarrow i+1}
=
\frac{N-i}{2 N},
T_{i \rightarrow i-1}
=
\frac{i}{2N},
$
%
describing a random walk with a drift 
towards the deterministic fixed point
$i=N/2$.
 In equilibrium, we have $P_i \, T_{i\rightarrow i+1} = P_{i+1} \, T_{i+1\rightarrow i}$
for every $i$,
thus
\begin{eqnarray}
P_{i+1} = P_0 \prod_{j=0}^{i} \frac{N-j}{j+1} =  P_0 \left( \begin{array}{c} N \\ i+1 \end{array}
\right).
\end{eqnarray}
$P_i$ is a binomial distribution 
around the equilibrium of the replicator dynamics.

The {\sl Prisoner's Dilemma}
 has a 
Nash equilibrium for mutual defection, i.e.\ $i=0$.
At the borders, for $T_{0\to 1}$ and $T_{N\to N-1}$ 
we assume a small mutation rate $\mu$ \cite{Fud04}.
As $b=0$, also state $i=1$ is absorbing for $w=1$ (two cooperators are
needed to promote cooperation).
Thus a small mutation rate $\mu$
has to be assumed also for $T_{1{\to}2}$.
Alternatively
one could assume $w<1$.

\begin{table}[htbp]
\caption{The four payoff matrix cases considered in
Fig. \ref{figfluktuat}
under the assumption of a small mutation rate
(see text).
Without the frequency-dependence introduced by a game,
the cases of constant fitness and neutral evolution
known from mathematical genetics are recovered;
the distribution keeps localized 
(for low mutation rates, mutants stay rare).
For the analytical expressions of the
distributions see \cite{Cla05}.
The distributions follow approximately a stretched exponential with the
fit exponents $\gamma$ as given in the table.
\label{tablestretched}}
\mbox{}\\[-1.8ex]
\begin{center}
\begin{tabular}{|c|cc|c|}
\hline Payoff matrix & Distribution & $\approx\exp(-bx^{\gamma})$ & Nash
equilibrium (NE)? \\ \hline
\begin{tabular}{c}$a=b=c=d$\\Neutral evolution\end{tabular}
& $P_i \sim \frac{1}{i(N-i)}$ && 
{ 
drift $\to i=0$ and $\to i=N$ 
}
\\\hline
\begin{tabular}{c}$a=b<c=d$\\ constant fitness\end{tabular}
&$\simeq$ exponential & $\gamma=0.87$&  drift $\to i=0$ \\\hline
$P_{AC}=\left(\begin{array}{cc}0&1\\1&0
\end{array}\right)$
&$\simeq$ binomial& $\gamma=2.07$ &
$i=N/2$ { ``internal NE''}
\\\hline
$P_{PD}
=\left(\begin{array}{cc}3&0\\5&1
\end{array}\right)$
& & $\gamma=0.63$& 
$i=0$  { ``external NE''}
\\\hline
\end{tabular}
\end{center}
\end{table}

\begin{figure}[thbp]
\begin{center}
\includegraphics[height=90mm,angle=270]{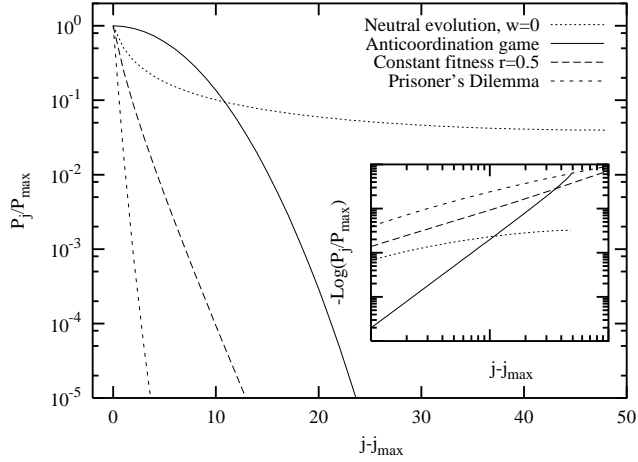}
\end{center}
\caption{Reproduced from \cite{Cla05}.
The invariant distributions $P_j$ share different decay tails
as a function of the distance 
$j-j_{\rm max}$ from the Nash equilibrium $j_{\rm max}$. Here $r=1-w+wa$.
In the four cases considered here
(shown for $N=100$), the distributions 
can be approximated by stretched exponentials 
(which would appear as straight lines in 
appropriate scaling, the inset shows that the
approximation is reasonable).
For higher mutation rates see
\cite{Tra06a}.
\label{figfluktuat}}
\end{figure}

The four cases are summarized in Table \ref{tablestretched}.
For {\sl neutral evolution} the decay is very slow. 
With {\sl constant fitness}, the probability decays approximately exponentially. 
For the Prisoner's dilemma, the decay becomes slower with larger distance, while it becomes faster for the binomial distribution at the internal Nash equilibrium.
The distributions are shown in Fig. \ref{figfluktuat}.
The decay of the distribution can be fitted by a
$P\approx\exp(-bx^\gamma)$,
$\gamma_{AC}=2.06$,
$\gamma_{CF}=0.87$,
$\gamma_{PD}=0.63$.
This corresponds to random motion in an anharmonic potential.

\section{From microscopic equations to macroscopic equations}
Corresponding to the discreteness of states, we first 
describe the processes by a master equation, then
formulate a Fokker-Planck (Kolmogorov forward) equation
for large $N$, and finally compare the 
deterministic equations 
resulting from the limit $N\to\infty$.

\subsection{Limit of large populations: From Master equation to 
Fokker-Planck equation.\label{sec_FPE}}
The equation of motion (or time evolution equation) 
for the stochastic process can be formulated in terms of the master equation
\begin{eqnarray}
 P^{\tau+1}(i) -  P^{\tau}(i)
&=&
\hphantom{+  ~}
P^{\tau}(i-1)  T^+(i-1) -P^{\tau}(i)  T^- (i)
\\
&&
+  ~  P^{\tau} (i+1)  T^- (i+1)
-P^{\tau}(i)  T^+ (i)
\nonumber
\end{eqnarray}
for the probability distribution,
i.e., $P^{\tau}(i)$ is the probability to be in state $i$ 
at time $\tau$.
For $N \gg 1$ we can proceed via a Kramers-Moyal expansion,
defining $x=i/N$ and $t=\tau/N$.
Then a formal Taylor expansion of $T$ and $\rho(x,t) = N\, P^{\tau}(i)$ 
yields, considering only the two leading terms of
the Taylor expansion, 
\begin{eqnarray}
\frac{d}{d t} \rho(x,t) &=& - \frac{d}{dx} \left[ a(x) \rho(x,t) \right] + \frac{1}{2} \frac{d^2}{dx^2} \left[ b^2(x)  \rho(x,t) \right]
\\
a(x)&=& 
T^+(x)-T^-(x)
\\
b(x) &=& 
\sqrt[]{\frac{1}{N} \left[T^+(x)+T^-(x) \right] }.
\end{eqnarray}
For large, but finite $N$, this equation has the form of a Fokker-Planck equation. 
This allows to generalize the diffusion approximation
(see e.g.\ \cite{Dro01}) to coevolution.
Since the internal noise is not correlated in time as subsequent update steps are independent, 
according to the It{\^o} calculus
a corresponding Langevin equation reads
\begin{eqnarray}
\label{langevin}
 \dot x = a(x) + b(x) \xi
\end{eqnarray}
where $\xi$ is uncorrelated Gaussian noise
and $b(x)=0$ for $x=0$ and $x=1$. 
We here see that the
noise 
is multiplicative and frequency-dependent.
\typeout{!!!!!!!!!!! Langevin !!!!!!!!!!!!!}

\subsection{Limit of infinite populations: 
Replicator equation and Adjusted replicator equation.
\label{secLIMIT}}
The leading order term does not vanish 
and describes the deterministic drift term.
Surprisingly, the replicator equation 
\cite{Tay78,Hof79,Zee81}
\begin{eqnarray}
 \dot{x}=
\frac{w
x (1-x) (\pi^A_x
-\pi^B_x
)
}{\Delta \pi_{\rm max}}
\end{eqnarray}
is obtained for the Local update process \cite{Tra05},
whereas the adjusted replicator equation 
\begin{eqnarray}
\dot{x}=
\frac{w
x (1-x) (\pi^A_x
-\pi^B_x
)
}{1-w + w\langle \pi(x) \rangle}
\end{eqnarray}
is the deterministic limit of the Moran process \cite{Tra05}.
For other processes, other differential equations
may be obtained, see Table \ref{table1}.

The difference between the two equations may be best illustrated
for the 
 Prisoner's Dilemma.
Here the 
(ordinary) replicator dynamics reads
$\dot x = - x (1-x^2)$,
whereas the adjusted replicator dynamics reads
$\dot x = \frac{- x (1-x^2) }{\frac{1-w}{w}+1+3 x -x^2}$.
In this case, the stability of fixed points is preserved,
as the additional denominator can be absorbed into a
dynamical rescaling of the time scale, commonly known
as velocity transform.
However, in asymmetric conflicts the change of time scale matters,
as will be detailed in the next section.

\section{Asymmetric and cyclic games.}
\label{Cla:Sec:Asym}
Cyclicity of evolutionary dynamics intuitively 
is contradictory to the
oversimplified picture of an absolute
{\sl fitness landscape}: 
any species having reached the maximal fitness value
within a population will outcompete all others, and
no cyclicity or oscillation can emerge.
However, in ecological competition, oscillations 
of populations are
quite generic, as known since Lotka \cite{Lot20} and
Volterra \cite{Vol26} 
(see also \cite{Mur04}).
Of special interest are cases where species cyclically
outcompete each other, as in 
Dawkins' caricature of mating strategies
(see Sections \ref{Cla:Sec:BOTS}--\ref{Cla:Sec:BOTSdrift}),
or the children's game 
{\sl Rock-Paper-Scissors},
where 
rock crushes scissors, 
paper covers rock,
and scissors cuts paper.
This situation has been spotted in the
territorial behaviour of lizards
\cite{Sin96,Zam00},
and also in E.coli bacteria
in vitro \cite{Ker02}
and in vivo \cite{Kir04}.
The Fokker-Planck equation,
in analogy to Sec.~\ref{sec_FPE},
for the Rock-Paper-Scissors
game has recently been given in
\cite{Rei06},
and generalizations to other cyclic 
evolutionary games are straightforward.
It is generally claimed that such cyclic coevolution
promotes biodiversity
\cite{Csa04,Cla08}.
However, its relevance in a more general picture of
``evolution at large'' remains an issue still to be investigated. 

\subsection{Asymmetric conflicts (bimatrix games).}
\label{Cla:Sec:BOTS}
As mentioned in the introduction, 
the payoff matrix for the opponent can be different 
from that earned by the first player.
To set this scenario into work, 
it is usually required that the game is played between
two disjunct populations (as below, female and male),
or that the interaction process itself is asymmetric,
e.g., that one player is ``initiative'' and gains
payoffs different from those earned in the
opponents' role.
Many social and economic situations bear such asymmetries,
but they are often too weak to be 
significantly extracted from data. 

An illustrative example of biological mating behaviour has
been given by Dawkins \cite{Daw76}.
Male and female each can occur with two genetic strategies,
``fast'' (male: philanderer, female: ``fast'')
and 
``slow'' (male: ``faithful'', female: ``coy'').
The payoff benefit of a child is assumed to be $b$ for both parents,
the total cost of raising an offspring is ($-2c$),
and the prolonged courtship that coy females insists on
add a burden of $a$ to both parents.
Coy females and male philanderers produce nothing and gain nothing.
The cost ($-2c$) is covered by both parents,
except for philanderer males and fast females,
where the female has to growup its offspring alone.
This translates into the payoff matrix
(with usual parameter choices
$a=3, b=15, c=10$,
see also Fig.\ \ref{fig_botsmatrix})
\begin{eqnarray}
\mbox{~~~~~~~~~~}
(\pi_{ij}^{\rm M},\pi_{ij}^{\rm F})=
\left(
\begin{array}{cc}
(b-c-a,b-c-a) & (b-c,b-c)\\
(0,0) & (b,b-2c)
\end{array}
\right)
= 
\left(
\begin{array}{cc}
(2,2) & (5,5)\\
(0,0) & (15,-5)
\end{array}
\right).
\end{eqnarray}
If the four conditions $a>0, c>0, b-c > 0, 2c-a-b>0$ are fulfilled,
the game is cyclic.
These conditions correspond to the arrows in Fig.\ \ref{fig_botsmatrix}.
The cyclicity of the game is preserved if we consider a normalized
version (being equivalent to the payoff matrix of ``Matching Pennies''
played by two players):
\begin{eqnarray}
(\pi_{ij}^{\rm M},\pi_{ij}^{\rm F})=
\left(
\begin{array}{cc}
(+1,-1) & (-1,+1)\\
(-1,+1) & (+1,-1)
\end{array}
\right).
\end{eqnarray}
For the cyclicity refer to Fig.~\ref{fig_botsmatrix}.
Let the cycle start in the lower right corner, where
fast females and male philandrers are present.
If males are
philanderers, it pays for females to be coy
(lower left corner);
 insisting on a long courtship period to make males invest more in the
offspring (upper left corner).
 However, once most males are faithful,
fast females are favored avoiding the costs of
courtship (upper right corner).
 Subsequently, the male investment
into the offspring is no longer justified, philanderers are again favored
(lower right corner), and the cycle continues.

\begin{figure}[htbp]
\begin{center}
\setlength{\unitlength}{1pt}
\begin{picture}(320,65)(-10,-10)
\put(0,5){\makebox(20,10)[r]{Philanderer}}
\put(0,25){\makebox(20,10)[r]{Faithful}}
\put(15,15){\makebox(20,10)[r]{Male}}
\put(55,45){\makebox(20,10){Coy}}
\put(105,45){\makebox(20,10){Fast}}
\put(80,55){\makebox(20,10){Female}}
\put(40,0){\line(1,0){100}}
\put(40,20){\line(1,0){100}}
\put(40,40){\line(1,0){100}}
\put(40,0){\line(0,1){40}}
\put(90,0){\line(0,1){40}}
\put(140,0){\line(0,1){40}}
\Red{\put(85,35){\vector(1,0){22.5}}
\put(100,15){\vector(-1,0){17.5}}
\put(105,20){\vector(0,-1){10}}
\put(55,7.5){\vector(0,1){10}}
}
\put(50,0){\makebox(10,10)[c]{$0$}}
\put(70,10){\makebox(10,10)[c]{$0$}}
\put(100,0){\makebox(10,10)[c]{$15$}}
\put(120,10){\makebox(10,10)[c]{$-5$}}
\put(50,20){\makebox(10,10)[c]{$2$}}
\put(70,30){\makebox(10,10)[c]{$2$}}
\put(100,20){\makebox(10,10)[c]{$5$}}
\put(120,30){\makebox(10,10)[c]{$5$}}

\put(185,5){\makebox(20,10)[r]{Philanderer}}
\put(185,25){\makebox(20,10)[r]{Faithful}}
\put(195,15){\makebox(20,10)[r]{Male}}
\put(230,45){\makebox(20,10)[c]{Coy}}
\put(280,45){\makebox(20,10)[c]{Fast}}
\put(255,55){\makebox(20,10)[c]{Female}}
\put(220,0){\line(1,0){100}}
\put(220,20){\line(1,0){100}}
\put(220,40){\line(1,0){100}}
\put(220,0){\line(0,1){40}}
\put(265,0){\line(0,1){40}}
\put(320,0){\line(0,1){40}}
\Red{
\put(265,35){\vector(1,0){22.5}}
\put(280,15){\vector(-1,0){17.5}}
\put(285,20){\vector(0,-1){10}}
\put(235,7.5){\vector(0,1){10}}
}
\put(230,0){\makebox(10,10)[c]{$-1$}}
\put(250,10){\makebox(10,10)[c]{$+1$}}
\put(280,0){\makebox(10,10)[c]{$+1$}}
\put(300,10){\makebox(10,10)[c]{$-1$}}
\put(230,20){\makebox(10,10)[c]{$+1$}}
\put(250,30){\makebox(10,10)[c]{$-1$}}
\put(280,20){\makebox(10,10)[c]{$-1$}}
\put(300,30){\makebox(10,10)[c]{$+1$}}
\end{picture}
\end{center}
\caption{Asymmetric payoff matrices for the Battle of the Sexes
\cite{Daw76}.
Left: Original payoff matrix suggested by Dawkins.
Right: Simplified payoff matrix, identical to the
game ``Matching pennies''.
In both cases, the left entry is for the row player, and the
right entry 
(upon a common convention,
shifted upwards, to enhance intuitive assignment)
is for the column player.
Arrows indicate the cyclic dominance.
\label{fig_botsmatrix}}
\end{figure}


This game exhibits 
neutrally stable periodic orbits
\cite{May87}
when described by the usual replicator equation approach
\cite{Tay78,Hof79,Zee81}.
This would, according to Dawkins, lead to an infinitely lasting oscillation
of strategies. 
As Dawkins argues, certain species 
(gibbon, stickleback, duck, fruit fly) 
have, in the course of real evolution, 
chosen to follow a fixed pair of strategies;
for them the ``Battle of the Sexes''
has come to a rest. 
Of course this models relies on several assumptions,
and in the remainder we will analyze how the
conclusions change when considering 
different evolutionary processes 
and when explicitly considering the
finiteness of the population.

\begin{figure}[htbp]
\begin{center}
\includegraphics[scale=0.80]{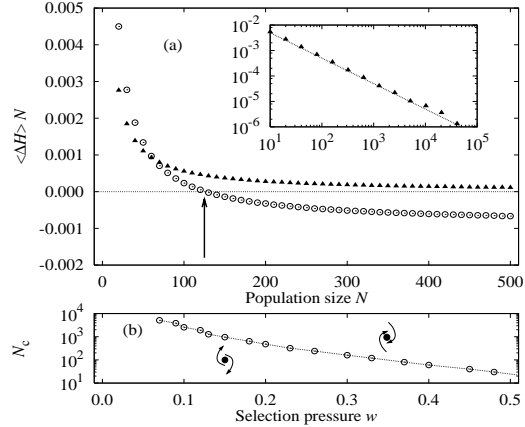}
\end{center}
\caption{Reproduced from \cite{Tra05}.
Drift reversal in the asymmetric game ``Battle of the Sexes''.
Shown is the average change of $H$ (which is a constant of motion
in the ordinary replicator equation) for different population
sizes.
(a)
For the Moran process (circles) above a critical population size 
the average change of $H$ becomes negative so that 
trajectories spiral outwards on average.
For the Local update, no drift reversal is observed
and for $N\to\infty$ the change of $H$ approaches
zero (shown in the inset in double logarithmic plot).
(b): Scaling of the critical population size with
selection pressure $w$.
\label{figreversal}}
\end{figure}

\subsection{Counterintuitive behaviour for the ``Battle of the Sexes''.}
\label{Cla:Sec:BOTSdrift}
The case of asymmetric conflicts opens the possibility of
a conterintuitive finite-size effect:
The sign of the average of $H=-xy(1-x)(1-y)$ 
(being a constant of motion for the ordinary replicator equation)
changes for
the
Moran
process
above a critical population size,
whereas the change remains positive (spiraling outwards) for the
Local update and Fermi process 
in all finite populations
(Fig.~\ref{figreversal})
\cite{Tra05,Cla06,Cla07}.
Hence in the $N\to\infty$ limit the Moran process 
shows a deterministic behaviour 
(that of the adjusted replicator equation),
which is qualitatively different from that of the
ordinary replicator equation.
The reason is that the common velocity transformation,
which can absorb the normalization denominator of the
average fitness, 
here cannot be performed, as female and male 
population earn different payoffs
and the denominators no longer coincide.
The adjusted replicator equations here read
\begin{eqnarray}
\dot x &=& + 2 \frac{(x^2-x) (2y-1)}{\frac{1-w}{w}+ (2y-1)(2x-1)} \\
\dot y &=& - 2 \frac{ (y^2-y) (2x-1)}{\frac{1-w}{w}- (2y-1)(2x-1)}.
\end{eqnarray}
Hence, if the two populations earn different average 
payoffs (here they differ by a factor $-1$), 
the fixed point stability of both types of 
replicator equations can differ.

\normalsize




%
 \normalsize 
\clearpage
\begin{thebibliography}{Nov0000} 
\small
\bibitem[Alb04]{Alb04} J.\ Alboszta and J.\ Mi\c{e}kisz, {\sl Stability of evolutionarily stable strategies in discrete replicator dynamics with time delay}, J.\ Theor.\ Biol.\  231 (2004), \href{http://dx.doi.org/doi:10.1016/j.jtbi.2004.06.012}{175-179}. \\[-4.5ex]
\bibitem[Ant06]{Ant06} T.\ Antal, S.\ Redner, and V.\ Sood, {\sl Evolutionary Dynamics on Degree-Heterogeneous Graphs},  Phys.\ Rev.\ Lett.\  96 (2006),  \href{http://dx.doi.org/10.1103/PhysRevLett.96.188104} {188104}. \\[-4.5ex]
\bibitem[Axe84]{Axe84} R.\ Axelrod, {\sl The evolution of cooperation}, Basic Books, New York (1984). \\[-4.5ex]
\bibitem[Baa97]{Baa97} E.\ Baake, M.\ Baake, and H.\ Wagner, {\sl Ising Quantum Chain is Equivalent to a Model of Biological Evolution}, Phys.\ Rev.\ Lett.\  78 (1997), \href{http://dx.doi.org/10.1103/PhysRevLett.78.559} {559-562}. \\[-4.5ex]
\bibitem[Bak93]{Bak93} P.\ Bak and K.\ Sneppen, {\sl Punctuated equilibrium and criticality in a simple model of evolution}, Phys.\ Rev.\ Lett.\  71 (1993), \href{http://dx.doi.org/10.1103/PhysRevLett.71.4083} {4083-4086}. \\[-4.5ex]
\bibitem[Ban05]{Ban05} C.\ Bandt, {\sl The Discrete Evolution Model of Bak and Sneppen is Conjugate to the Classical Contact Process},  J.\ Stat.\ Phys.\ 120 (2005), \href{http://dx.doi.org/10.1007/s10955-005-5965-x} {685-693}. \\[-4.5ex]
\bibitem[Blu93]{Blu93} L.\ E.\ Blume,  {\sl The Statistical Mechanics of Strategic Interaction},  Games Econom.\ Behav.\  5 (1993), 387-424. \\[-4.5ex]
\bibitem[Che03]{Che03} J.\ L.\ Cherry and J.\ Wakely, {\sl A diffusion approximation for selection and drift in a subdivided population},  Genetics 163 (2003), 421-428. \\[-4.5ex]
\bibitem[Cla05]{Cla05} J.\ C.\ Claussen and A.\ Traulsen, {\sl  Nongaussian fluctuations arising from finite populations: Exact results for the evolutionary Moran process},  Phys.\ Rev.\ E 71 (2005), \href{http://dx.doi.org/10.1103/PhysRevE.71.025101} {025101(R)}.  \\[-4.5ex]
\bibitem[Cla06]{Cla06} J.\ C.\ Claussen and A.\ Traulsen, {\sl Fluctuations in Coevolutionary Dynamics and Implications for Multi-Agent Models}, p.\ 411-419, in: D.\  Helbing (Ed.), Proc.\ Potentials of Complexity Science for Business, Government, and the Media, Budapest (2006). \\[-4.5ex]
\bibitem[Cla07]{Cla07} J.\ C.\ Claussen, {\sl Drift reversal in asymmetric coevolutionary conflicts: Influence of microscopic processes and population size},  Eur.\ Phys.\ J.\ B {\bf 60},  \href{http://dx.doi.org/10.1140/epjb/e2007-00357-2}{391-399} (2007). \\[-4.5ex]
\bibitem[Cla08]{Cla08} J.\ C.\ Claussen and A.\ Traulsen, {\sl Cyclic Dominance and Biodiversity in Well-Mixed Populations}, 
Phys.\ Rev.\ Lett. {\bf 100}, (2008) \href{http://dx.doi.org/10.1103/PhysRevLett.100.058104}{058104}. \\[-4.5ex]
\bibitem[Csa02]{Csa04} T.\ L. Cz\'ar\'an, R.\ F.\ Hoekstra and L.\ Pagie, {\sl Chemical warfare between microbes promotes biodiversity}, Proc.\ Natl.\ Acad.\ Sci.\  USA 99 (2002),  \href{http://dx.doi.org/10.1073/pnas.012399899}{786-790}.\\[-4.5ex]
\bibitem[Daw76]{Daw76} R.\ Dawkins, {\sl The Selfish Gene}, Oxford University Press, New York (1976). \\[-4.5ex]
\bibitem[Dro01]{Dro01} B.\ Drossel, {\sl Biological evolution and statistical physics},  Adv.\ Phys.\  50 (2001), \href{http://dx.doi.org/10.1080/00018730110041365}{209-295}. \\[-4.5ex]
\bibitem[Ebe02]{Ebe02} H.\ Ebel and S.\ Bornholdt, {\sl Coevolutionary games on networks}, Phys.\ Rev.\ E 66 (2002), \href{http://dx.doi.org/10.1103/PhysRevE.66.056118} {056118}. \\[-4.5ex]
\bibitem[Eld72]{Eld72} N.\ Eldredge and S.\ J.\ Gould, {\sl Punctuated equilibria: an alternative to phyletic gradualism},  in: T.J.M. Schopf, Editor, Models in Paleobiology, Freeman Cooper, San Francisco, pp. 82-115 (1972). \\[-4.5ex]
\bibitem[Fis30]{Fis30} R.\ A.\ Fisher, {\sl The Genetical Theory of Natural Selection} (Clarendon, Oxford, 1930). \\[-4.5ex]
\bibitem[Fud04]{Fud04} D.\ Fudenberg and L.\ A.\ Imhof, {\sl Imitation Processes with Small Mutations}, Harvard Institute of Economic Research Discussion Paper No. 2050 (2004). \\[-4.5ex]
\bibitem[H\aa{}r02]{Har02} K.\ C.\ Harding and J.\ M.\ McNamara, {\sl A Unifying Framework for Metapopulation Dynamics}, Amer.\ Naturalist,  160 (2002), \href{http://dx.doi.org/10.1086/341014} {173-185}. \\[-4.5ex]
\bibitem[Hau05]{Hau05} C.\ Hauert and G.\ Szab\'{o}, {\sl Game theory and physics},  Am.\ J.\ Phys.\ 73 (2005), 405-414. \\[-4.5ex]
\bibitem[Hel92a]{Hel92a} D.\  Helbing, {\sl A Mathematical Model for Attitude Formation by Pair Interactions}, Behavioral Science  37 (1992), 190-214. \\[-4.5ex]
\bibitem[Hel92b]{Hel92b} D.\  Helbing,  {Phys.\  A   181(1992), 29-52};  {Phys.\  A   193 (1993), 241-258}; {Phys.\  A  196 (1993), 546-573}. \\[-4.5ex]
\bibitem[Hel96]{Hel96} D.\ Helbing, {\sl A stochastic behavioral model and a ``microscopic'' foundation of evolutionary game theory},  Theory and Decision  40 (1996), 149-179. \\[-4.5ex] 
\bibitem[Her94]{Her94} A.\ V.\ M.\ Herz, {\sl Collective Phenomena in Spatially Extended Evolutionary Games}, Journal of Theoretical Biology  169 (1994),  \href{http://dx.doi.org/doi:10.1006/jtbi.1994.1130}{65-87}. \\[-4.5ex]
\bibitem[Hof79]{Hof79} J.\ Hofbauer, P.\ Schuster, and K.\ Sigmund, {\sl A note on evolutionary stable strategies and game dynamics},  J.\ Theor.\ Biol.\  81 (1979), 609-612. \\[-4.5ex]
\bibitem[Hof84]{Hof84} J.\ Hofbauer and K.\ Sigmund, {\sl Evolutionstheorie und dynamische Systeme}, Paul Parey, Berlin (1984). \\[-4.5ex]
\bibitem[Hof98]{Hof98} J.\ Hofbauer  and K.\ Sigmund 1998, {\sl  Evolutionary Games and Population Dynamics} (Cambridge: Cambridge University Press) \\[-4.5ex]
\bibitem[Hof00]{Hof00} J.\ Hofbauer, K.\ H.\ Schlag, {\sl Sophisticated imitation in cyclic games}, J.\ Evol.\ Econ.\  10 (2000), \href{http://dx.doi.org/10.1007/s001910000049}{523-543}. \\[-4.5ex]
\bibitem[Jen04]{Jen04} H.\ J.\ Jensen, {\sl Emergence of species and punctuated equilibrium in the Tangle Nature model of biological evolution}, Phys.\  A  340 (2004), \href{http://dx.doi.org/doi:10.1016/j.physa.2004.05.022} {697-704}. \\[-4.5ex]
\bibitem[Ker02]{Ker02} B.\ Kerr, M.\ A.\ Riley, Marcus W.\ Feldman, and Brendan J.\ M.\ Bohannan, {\sl Local dispersal promotes biodiversity in a real-life game of rock-paper-scissors}, Nature   418 (2002), 171-174. \\[-4.5ex]
\bibitem[Kir04]{Kir04} B.\ C.\ Kirkup and M.\ A.\ Riley, {\sl Antibiotic-mediated antagonism leads to a bacterial game of  rock-paper-scissors in vivo}, Nature    428 (2004), 412. \\[-4.5ex]
\bibitem[Lie05]{Lie05} E.\ Lieberman, C.\ Hauert, M.\ A.\  Nowak,  {\sl Evolutionary Dynamics on Graphs},  Nature    433 (2005), 312-316. \\[-4.5ex]
\bibitem[Lin94]{Lin94} K.\ Lindgren and M.\ Nordahl,  {\sl Evolutionary dynamics of spatial games}, Phys.\  D    75 (1994), 292. \\[-4.5ex]
\bibitem[Lot20]{Lot20} A.\ J.\ Lotka, J.\ Am.\ Chem.\ Soc.\  42 (1920),  \href{http://dx.doi.org/10.1021/ja01453a010} {1595-1599}. \\[-4.5ex]
\bibitem[May73]{May73} J.\ Maynard Smith and G.\ Price, {\sl The logic of animal conflict}, Nature   246 (1973), 15. \\[-4.5ex]
\bibitem[May87]{May87} J.\ Maynard Smith and J.\ Hofbauer,  {\sl The ``battle of the sexes'': A genetic model with limit cycle behavior}, Theoret.\ Population Biol.\  32 (1987), 1-14. \\[-4.5ex]

\bibitem[Mie07]{Mie07} J.\ Miekisz, {\sl Evolutionary game theory and population dynamics} (2007), arXiv:q-bio/\href{http://arxiv.org/abs/q-bio/0703062}{0703062}, Lecture Notes in Math.\ (in press).

\bibitem[Mor62]{Mor62} P.\ A.\ P.\ Moran, {\sl The Statistical Processes of Evolutionary Theory}, (Clarendon, Oxford, 1962). \\[-4.5ex]
\bibitem[Mur04]{Mur04} J.\ D.\ Murray, {\sl Mathematical Biology I}, Springer (2004). \\[-4.5ex]
\bibitem[Nas51]{Nas51} J.\ Nash, {\sl Non-cooperative games}, {Ann.\ Math.\   54 (1951), 287-295}. \\[-4.5ex]
\bibitem[Nei04]{Nei04} D.\ B.\ Neill, {\sl Evolutionary stability for large populations},  J.\ Theor.\ Biol.\   227 (2004), 397-401. \\[-4.5ex]
\bibitem[Neu53]{Neu53} J.\ von Neumann and O.\ Morgenstern {\sl Theory of games and economic behavior}, Princeton University Press (1953). \\[-4.5ex]
\bibitem[Now92]{Now92} M.\ A.\ Nowak and R.\ M.\ May,  {\sl Evolutionary games and spatial chaos}, Nature  395 (1992), 826. \\[-4.5ex]
\bibitem[Now04]{Now04} M.\ A.\ Nowak, A.\ Sasaki, C.\ Taylor, and D.\ Fudenberg,  {\sl Emergence of cooperation and evolutionary stability in finite populations}, Nature   428 (2004), 646-650. \\[-4.5ex]
\bibitem[Oht06]{Oht06} H.\ Ohtsuki, C.\ Hauert, E.\ Lieberman, and M.\ Nowak, {\sl  A simple rule for the evolution of cooperation on graphs},  Nature  441 (2006), 502-505. \\[-4.5ex]
\bibitem[Pis97]{Pis97} Yu.\ M.\ Pis'mak, {\sl Solution of the master equation for the Bak-Sneppen model of biological evolution in a finite ecosystem}, Phys.\ Rev.\ E  56 (1997), \href{http://dx.doi.org/10.1103/PhysRevE.56.R1326}{1326-1329}. \\[-4.5ex]
\bibitem[Rei06]{Rei06} T.\ Reichenbach, M.\ Mobilia, and E.\ Frey, {\sl Coexistence versus extinction in the stochastic cyclic Lotka-Volterra model}, Phys.\ Rev.\ E  74 (2006), \href{http://dx.doi.org/10.1103/PhysRevE.74.051907}{051907}.  \\[-4.5ex]
\bibitem[Sin96]{Sin96} B.\ Sinervo and C.\ M.\ Lively, {\sl The rock-paper-scissors game and the evolution of alternative male strategies}, Nature   380 (1996), 240-243. \\[-4.5ex]
\bibitem[Sza98]{Sza98} G.\ Szab\'{o} and C.\ T\H{o}ke, {\sl Evolutionary prisoner’s dilemma game on a square lattice}, Phys.\ Rev.\ E   58 (1998), \href{http://dx.doi.org/10.1103/PhysRevE.58.69}{69-73}. \\[-4.5ex]
\bibitem[Sza99]{Sza99} G.\ Szab\'{o}, M.\ A.\ Santos, and J.\ F.\ F.\ Mendes, {\sl Vortex dynamics in a three-state model under cyclic dominance}, Phys.\ Rev.\ E   60 (1999), \href{http://dx.doi.org/10.1103/PhysRevE.60.3776} {3776-3780}. \\[-4.5ex]
\bibitem[Sza04]{Sza04b} {G.\  Szab\'{o}, A.\  Szolnoki, and R.\ Izs\'{a}k,} {\sl Rock-scis\-sors-pa\-per game on reg- ular small-world networks}, Journal of Physics A  37 (2004), \href{http://dx.doi.org/10.1088/0305-4470/37/7/006} {2599}. \\[-4.5ex]
\bibitem[Sza05]{Sza05} G.\  Szab\'{o}, J.\ Vukov, and A.\  Szolnoki, {\sl Phase diagrams for an evolutionary prisoner's dilemma game on two-dimensional lattices}, Phys.\ Rev.\ E  72 (2005), \href{http://dx.doi.org/10.1103/PhysRevE.72.047107} {047107}. \\[-4.5ex]
\bibitem[Sza07]{Sza07} G.\  Szab\'{o} and G.\ Fath, {\sl Evolutionary games on graphs}, Phys.\ Rep.\ 446 (2007), \href{http://dx.doi.org/10.1016/j.physrep.2007.04.004} {97-216}. \\[-4.5ex]
\bibitem[Szo04]{Sza04a} A.\  Szolnoki and G.\  Szab\'{o}, {\sl Vertex dynamics during domain growth in three-state models}, Phys.\ Rev.\ E   70 (2004), \href{http://dx.doi.org/10.1103/PhysRevE.70.027101}{027101}. \\[-4.5ex]
\bibitem[Szn00]{Szn00} K.\ Sznajd-Weron and J.\ Sznajd, {\sl Opinion evolution in closed community}, Internat.\ J Modern Phys.\ C  11 (2000), \href{http://dx.doi.org/10.1142/S0129183100000936}{1157-1165}. \\[-4.5ex]
\bibitem[Tay78]{Tay78} P.\ D.\ Taylor and L.\ B.\ Jonker, {\sl Evolutionary stable strategies and game dynamics},  Math.\ Biosci.\ 40 (1978), 145-156. \\[-4.5ex]
\bibitem[Tay04]{Tay04} C.\ Taylor, D.\ Fudenberg, A.\ Sasaki, and M.\ A.\ Nowak,  {\sl Evolutionary game dynamics in finite populations}, Bull.\ Math.\ Biol.\  66 (2004), 1621-1644. \\[-4.5ex]
\bibitem[Tay06]{Tay06} C.\ Taylor and M.\ A.\ Nowak, {\sl Evolutionary game dynamics with non-uniform interaction rates}, Theoret.\ Population Biol.\ 69 (2006), 243-252. \\[-4.5ex]
\bibitem[Tra04]{Tra04} A.\ Traulsen and J.\ C.\ Claussen, Similarity based cooperation and spatial segregation,  Phys.\ Rev.\ E  70 (2004), \href{http://dx.doi.org/10.1103/PhysRevE.70.046128} {046128}. \\[-4.5ex]
\bibitem[Tra05]{Tra05} A.\ Traulsen, J.\ C.\ Claussen, and C.\ Hauert, {\sl Coevolutionary Dynamics: From Finite to Infinite Populations}, Phys.\ Rev.\ Lett.\  95 (2005), \href{http://dx.doi.org/10.1103/PhysRevLett.95.238701} {238701}. \\[-4.5ex]
\bibitem[Tra06a]{Tra06a} A.\ Traulsen, J.\ C.\ Claussen, and C.\ Hauert, {\sl Coevolutionary dynamics in large, but finite populations,} Phys.\ Rev.\  E 74 (2006), \href{http://dx.doi.org/10.1103/PhysRevE.74.011901} {011901}. \\[-4.5ex]
\bibitem[Tra06c]{Tra06c} A.\ Traulsen, M.\ A.\ Nowak and J.\ Pacheco, {\sl Stochastic dynamics of invasion and fixation}, Phys.\ Rev.\ E  74 (2006), \href{http://dx.doi.org/10.1103/PhysRevE.74.011909} {011909}. \\[-4.5ex]
\bibitem[Tra06b]{Tra06b} A.\ Traulsen, J.\ M. Pacheco, and L.\ A.\ Imhof, {\sl Stochasticity and evolutionary stability}, Phys.\ Rev.\ E  74 (2006), \href{http://dx.doi.org/10.1103/PhysRevE.74.021905} {021905}. \\[-4.5ex]
\bibitem[Tra07]{Tra07} A.\ Traulsen, Martin A.\ Nowak and J.\ Pacheco,  Stochastic payoff evaluation increases the temperature of selection, J.\ Theor.\ Biol.\  244 (2007), 349-356. \\[-4.5ex]
\bibitem[Vol26]{Vol26} V.\ Volterra, Mem.\ Accad.\ Lincei.\  2 (1926), 31, translation in: {Animal Ecology},  R.\ N.\ Chapman (McGraw Hill, 1931), 409-448. \\[-4.5ex]
\bibitem[Wil04]{Wil04} G.\ Wild and P.\ D.\ Taylor,  {\sl Fitness and evolutionary stability in game theoretic models of finite populations}, Proc R.\ Soc.\ Lond.\ Ser.\ B Biol.\ Sci.\  271 (2004), 2345-2349. \\[-4.5ex]
\bibitem[Wri31]{Wri31} S.\ Wright, {\sl Evolution in Mendelian Populations}, \href{http://www.genetics.org/content/vol16/issue2/}{Genetics 16} (1931), 97. \\[-4.5ex]
\bibitem[Zam00]{Zam00} K.\ R.\ Zamudio and B.\ Sinervo,  {\sl Polygyny, mate-guarding, and posthumous fertilization as alternative male mating strategies}, Proc.\ Natl.\ Acad.\ Sci.\  USA 97 (2000),   \href{http://dx.doi.org/10.1073/pnas.011544998} {14427}. \\[-4.5ex]
\bibitem[Zee80]{Zee80} E.\ C.\ Zeeman, {\sl Population dynamics from game theory}, in: A.\ Nitecki, C.\ Robinson (Eds.),  Proceedings of an International Conference on Global Theory of Dynamical Systems,  Lecture Notes in Math.\  819, Springer, Berlin, 1980. \\[-4.5ex]
\bibitem[Zee81]{Zee81} E. Zeeman, {\sl Dynamics of the evolution of animal conflicts}, J.\ Theor.\ Biol.\   89 (1981), 249-270. \\[-4.5ex] 








%
\end{thebibliography}
\end{document}